\documentclass[twocolumn,prl,showpacs,showkeys,amsmath,amssymb]{revtex4}
\usepackage{graphicx,amsmath,epsfig,latexsym,color}
\newcommand{\be}{\begin{equation}}
\newcommand{\ee}{\end{equation}}
\begin{document}
\title{Measurement of atomic diffraction phases induced by material gratings}
\author{John D.\ Perreault and Alexander D.\ Cronin}
\affiliation{University of Arizona, Tucson, Arizona 85721}
\date{\today}
\begin{abstract}
Atom-surface interactions can significantly modify the intensity
and phase of atom de Broglie waves diffracted by a silicon nitride
grating. This affects the operation of a material grating as a
coherent beam splitter. The phase shift induced by diffraction is
measured by comparing the relative phases of serveral interfering
paths in a Mach-Zehnder Na atom interferometer formed by three
material gratings.  The values of the diffraction phases are
consistent with a simple model which includes a van der Waals
atom-surface interaction between the Na atoms and the silicon
nitride grating bars.
\end{abstract}
 \maketitle

A coherent beam splitter is a useful component for constructing an
atom interferometer \cite{berm97}.  The purpose of the beam
splitter is to generate a quantum superposition of atom waves,
propagating along two paths which can be recombined to form an
interference pattern. The contrast and phase of the interference
pattern can then be used to study interactions that affect the
atoms differently in the two interferometer paths.  However, atom
beam splitters formed using laser light
\cite{weit94,durr98,buch03} and material grating structures
\cite{perr05} can create beams with differing complex amplitudes.
A familiar analogy of this in optics occurs for light beam
splitters formed using glass plates, thin metal films, and
multi-layer dielectric stacks which all cause a phase shift
between the reflected and transmitted components \cite{born99}.
These complex amplitudes are an important concern when building an
atom interferometer since they affect the phase and contrast of
the interference pattern, and have been identified as a source of
uncertainty for atom interferometer gyroscopes \cite{gust00}. Here
we present the first evidence of beam splitter induced phase
shifts in an atom interferometer based on material gratings and a
new method for measuring these phase shifts for interferometers
that utilize diffraction.

The van der Waals (vdw) atom-surface interaction \cite{milo94}
plays a significant role in determining the intensity and phase
(diffraction phase) of atom waves split by a material grating.
Several atom-optics experiments have observed how atom-surface
interactions can affect the intensity of atom waves
\cite{gris99,cron04,perr05}. By comparison, few experiments have
directly measured the diffraction phases induced by atom-surface
interactions \cite{pifm05}.  In this Letter the first and second
order diffraction phases are measured by comparing the phase
difference between the various interfering paths in a three
grating Mach-Zehnder atom interferometer.

\begin{figure}
\scalebox{.65}{\includegraphics{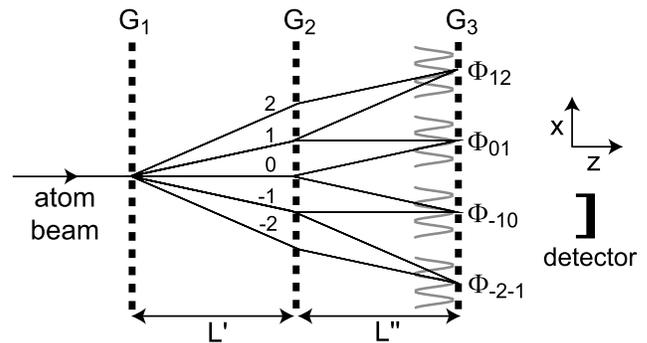}}
\caption{\label{fig:setup}Diagram of experimental setup for
measuring diffraction phases with an atom interferometer.  Since
the interference pattern is read out using $G_{3}$ as a mask, only
the paths indicated by the solid lines will lead to a significant
interference signal.  The mismatch of the grating spacings $\Delta
L\equiv L'' - L'$ and the diffraction phases $\Phi_{n}$ induced by
$G_{1}$ determine the measured interference phases $\Phi_{mn}$.}
\end{figure}

The experimental setup used to measure the atomic diffraction
phases induced by a material grating is shown in Fig.
\ref{fig:setup}.  A Mach-Zehnder atom interferometer, similar to
the one described in \cite{keit91}, is formed using zeroth and
first order diffraction from three 100 nm period silicon nitride
gratings \cite{sava96} which are denoted $G_{1}$,$G_{2}$,$G_{3}$.
These gratings are nominally separated by 1 m.  A collimated
supersonic Na atom beam is first diffracted by grating $G_{1}$,
inducing a phase shift which depends on the diffraction order and
will be described later. Each diffracted beam then undergoes first
order diffraction by $G_{2}$ and forms a spatial interference
pattern just before $G_{3}$. The atoms transmitted through $G_{3}$
are ionized by a $60\ \mu$m wide hot Re wire and counted by a
channel electron multiplier.  Grating $G_{3}$ is then scanned in
the direction transverse to the incident atom beam (x-axis) to
determine the phase of the interference pattern. While there are
many paths which can interfere at the plane of $G_{3}$ only the
ones which involve first order diffraction by $G_{2}$ (as
indicated in Fig. \ref{fig:setup}) will lead to a significant
interference signal because of velocity dispersion and the use of
$G_{3}$ as a transmission mask.  Since each relevant path through
$G_{2}$ undergoes first order diffraction both paths acquire the
same diffraction phase, which means there is no net phase shift
induced by $G_{2}$. In addition, grating $G_{3}$ acts as a
transmission mask so only the diffraction phases induced by
$G_{1}$ will lead to a relative phase shift between the
interfering paths.  In principle, the diffraction phases can then
be determined by comparing the phase of the various interferometer
outputs which can be measured separately by moving the detector
along the x-axis.

In practice there are two types of phase shifts that need to be
considered when predicting the relative phase of the various
interferometer outputs.  One originates from the diffraction phase
induced by $G_{1}$ and the other from a distance mismatch between
the gratings $G_{1}$,$G_{2}$,$G_{3}$, which is denoted by $\Delta
L\equiv L'' - L'$ in Fig. \ref{fig:setup}. In order to report a
measurement of the diffraction phases, expressions for both of
these phase shifts will be put forth.

\begin{figure}
\scalebox{.45}{\includegraphics{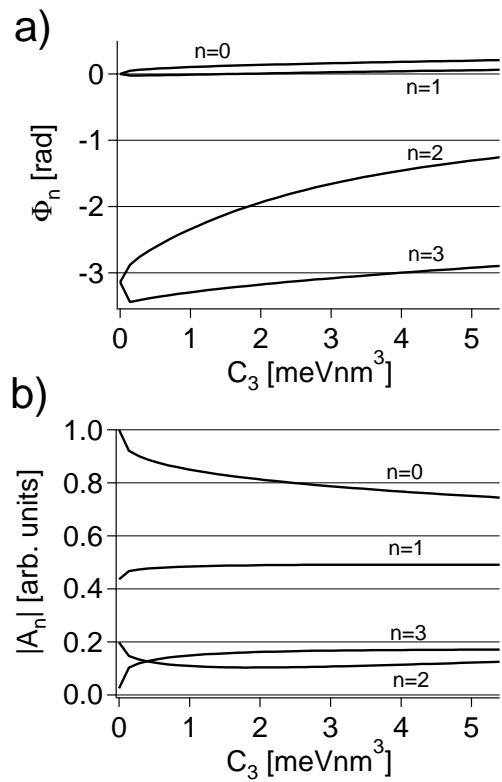}}
\caption{\label{fig:theory}Plots of the diffraction phases
$\Phi_{n}$ and amplitudes $A_{n}$ as a function of the vdW
coefficient $C_{3}$ for diffraction orders $n=0,1,2,3$ according
to Eqn. \ref{eq:psi}. The grating parameters $w=65$ nm, $l=150$
nm, $d=100$ nm, and atom beam velocity $v=2900$ m/s were used to
generate the curves.}
\end{figure}

From previous work it has been shown that atomic diffraction from
a material grating will create diffracted beams with complex
amplitudes $\psi_{n}$ given by\be
\begin{split}
\psi_{n}=A_{n}e^{i\Phi_{n}}\propto\int_{-w/2}^{w/2}e^{i\phi(\xi)}e^{i2\pi\xi
n/d}d\xi,
\end{split}
\label{eq:psi}\ee where $A_{n}$ is the amplitude and $\Phi_{n}$ is
the diffraction phase for a given diffraction order $n$ as derived
in \cite{cron04,perr05,pifm05}.  The variable $\xi$ is the
position measured from the center of the grating window whose size
is $w$ and grating period is $d$. The expression in Eqn.
\ref{eq:psi} is valid in the far-field (Fraunhofer) diffraction
regime and is appropriate for our experimental setup as described
in \cite{perr05}.  The phase $\phi(\xi)$ represents the phase
accumulated by the atom wave as it propagates through the grating
window, given by the WKB approximation to leading order in
$V(\xi)$ as\be
\begin{split}
\phi(\xi) = -\frac{lV(\xi)}{\hbar v},
\end{split}
\label{eq:phixi}\ee where $l$ is the thickness of the grating,
$\hbar$ is Planck's constant and $v$ is the atom beam velocity
\cite{perr05}. The atom-surface interaction potential $V(\xi)$ in
Eqn. \ref{eq:phixi} is given by\be
\begin{split}
V(\xi)=-C_{3}\left[\left|\xi-\frac{w}{2}\right|^{-3} +
\left|\xi+\frac{w}{2}\right|^{-3}\right],
\end{split}
\label{eq:potential}\ee where $C_{3}$ is coefficient describing
the strength of the vdW interaction.  This form of the vdW
potential is valid for atom-surface distances of $\lesssim 1\mu$m
for Na atoms and neglects the finite thickness of the grating bars
\cite{milo94,perr05}. A plot of diffraction phases and amplitudes
are shown in Fig. \ref{fig:theory} as a function of $C_{3}$.

There is also a phase shift between the interferometer outputs
induced by a distance mismatch of the gratings
$G_{1}$,$G_{2}$,$G_{3}$ along the z-axis.  The origin of this
phase shift can be understood by recalling that when two plane
waves interfere at an angle $\theta=\lambda_{dB}d^{-1}$,
interference fringes will be formed with intensity maxima along
lines with an angle $\theta/2$ as described in \cite{hech98}.  In
other words, a spatial interference pattern of the form
$\cos[k_{g}(x - z\theta/2)]$ with wave number $k_{g}=2\pi/d$ will
be observed.  If the $n=0,1$ paths depicted in Fig.
\ref{fig:setup} are regarded as plane waves interfering at an
angle $\theta=\lambda_{dB}d^{-1}$ then by geometry an effective
phase shift\be
\begin{split}
\Phi_{\Delta L} = \frac{\pi\Delta L \lambda_{dB}}{d^{2}},
\end{split}
\label{eq:phidl}\ee will be observed if there is a distance
mismatch $\Delta L\equiv L'' - L'$ between the interferometer
gratings. For two general paths originating from the diffraction
orders $m,n$ of $G_{1}$ the effective phase shift induced by
$\Delta L$ will be given by $(m+n)\Phi_{\Delta L}$, since the
fringe maxima are just rotated by an angle $(m+n)\theta/2$ with
respect to the z-axis.

The expressions for the diffraction phase $\Phi_{n}$ (Eqn.
\ref{eq:psi}) and grating distance mismatch phase $\Phi_{\Delta
L}$ (Eqn. \ref{eq:phidl}) can then be used to specify the wave
function $|\chi_{mn}\rangle$ just before grating $G_{3}$\be
\begin{split}
|\chi_{mn}\rangle=e^{ik_{g}x}|\psi_{m}\rangle +
e^{i(m+n)\Phi_{\Delta L}}|\psi_{n}\rangle,
\end{split}
\label{eq:chi}\ee for any two interfering paths involving the
diffraction orders $m,n$ of grating $G_{1}$.  The wave functions
$|\psi_{m}\rangle$ and $|\psi_{n}\rangle$ describe the two atom
beams corresponding to a given interferometer output and accounts
for the diffraction phases and amplitudes through the relations\be
\begin{split}
\langle\psi_{n}|\psi_{n}\rangle\equiv|A_{n}|^2,
\end{split}
\label{eq:amp}\ee and \be
\begin{split}
\langle\psi_{n}|\psi_{m}\rangle\equiv
A_{n}^{*}A_{m}e^{i(\Phi_{m}-\Phi_{n})},
\end{split}
\label{eq:phase}\ee where $A_{m},A_{n}$ and $\Phi_{m},\Phi_{n}$
are given by Eqn. \ref{eq:psi}.  The intensity can then be found
in the usual way
\begin{eqnarray}
I_{mn}(x) & \equiv & \langle\chi_{mn}|\chi_{mn}\rangle\nonumber\\
& \propto & 1 + C_{mn}\cos(k_{g}x + \Phi_{mn}),
\label{eq:intensity}\end{eqnarray} where $C_{mn}$ and $\Phi_{mn}$
are the observed contrast and phase for a given interferometer
output involving the diffraction orders $m,n$ as depicted in Fig.
\ref{fig:setup}.  The measured interferometer phase\be
\begin{split}
\Phi_{mn} \equiv \left(\Phi_{m} - \Phi_{n}\right) -
(m+n)\Phi_{\Delta L},
\end{split}
\label{eq:phiifm}\ee can also be expressed in terms of the
diffraction phases $\Phi_{n}$ and grating mismatch phase shift
$\Phi_{\Delta L}$. From Eqn. \ref{eq:phiifm} it can be seen that
$\Phi_{mn}=-\Phi_{-m-n}$, as implied by the symmetry of the
interferometer.

\begin{figure}
\scalebox{.67}{\includegraphics{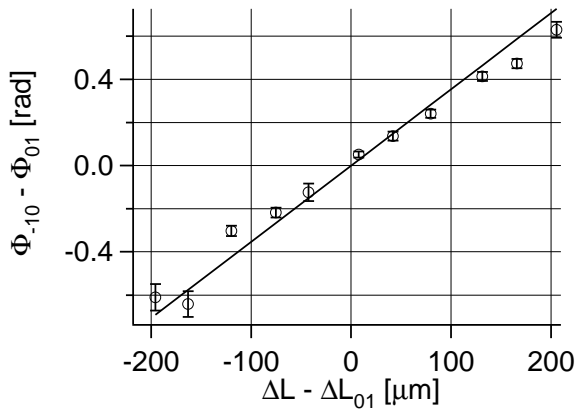}}
\caption{\label{fig:phivsdl}Measured phase difference as a
function of $\Delta L$.  The variable $\Delta L_{01}$ is the
grating distance mismatch required to make the observed phase
difference zero.  The solid curve contains no free parameters and
is the phase difference implied by Eqns. \ref{eq:phidl} and
\ref{eq:phiifm} for the independently measured velocity $v=2900$
m/s ($\lambda_{dB}=0.056$ \AA) and $d=100$ nm.}
\end{figure}

Equations \ref{eq:intensity} and \ref{eq:phiifm} can now be used
to predict the phase shift between the various interferometer
outputs shown in Fig \ref{fig:setup}.  One noteworthy aspect of
Eqn. \ref{eq:phiifm} is the possibility for $\Phi_{mn}$ to be made
zero if the diffraction phase term is cancelled by the appropriate
choice of $\Delta L$.  To verify this prediction the phase
difference $\Phi_{-10}-\Phi_{01}=-2\Phi_{01}$ was measured as a
function of $\Delta L$ and the results are shown in Fig.
\ref{fig:phivsdl}. The atom beam velocity $v=2900$ m/s
($\lambda_{dB}=0.056$ \AA) was measured independently by observing
the diffraction pattern generated by $G_{1}$ when the other
gratings are removed.  Equations \ref{eq:phidl} and
\ref{eq:phiifm} can then be used to generate the solid curve in
Fig. \ref{fig:phivsdl}, since the grating period is known to be
$d=100$ nm. The data agree quite well with our theory considering
that there are no free parameters in the solid curve.  At this
point $\Delta L$ is known only up to some offset $\Delta L_{01}$
which can be interpreted as the grating distance mismatch which
leads to the special case of $\Phi_{\Delta L}=\Phi_{0}-\Phi_{1}$.
However, the value of $\Delta L_{01}$ will be determined next
using another independent measurement technique.

\begin{figure}
\scalebox{.59}{\includegraphics{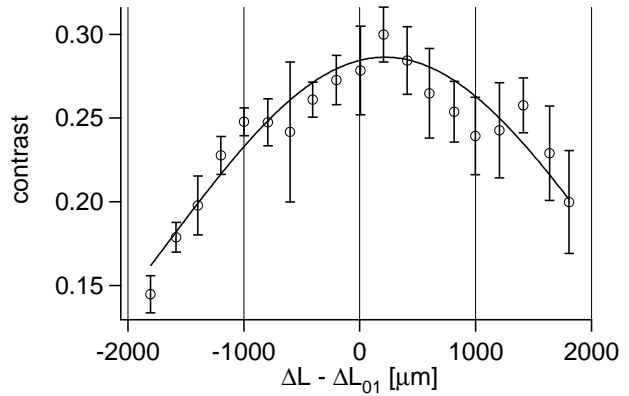}}
\caption{\label{fig:cvsdl}Measured contrast as a function of the
grating distance mismatch $\Delta L$.  The solid line is a best
fit using Eqn. \ref{eq:cspat}, which led to the determination of
$\Delta L_{01}=218\pm 72\ \mu$m.}
\end{figure}

As $\Delta L$ is shifted further from zero the partial spatial and
temporal coherence of the atom beam causes the measured
interference contrast to decrease.  This notion is evident in the
experimental data presented in Fig. \ref{fig:cvsdl}, which plots
the observed interference contrast as a function of $\Delta L$.
This contrast peak can be used to determine the offset $\Delta
L_{01}$, which then specifies the value of $\Phi_{0}-\Phi_{1}$
through Eqns. \ref{eq:phidl} and \ref{eq:phiifm}.  For our
particular atom beam setup spatial coherence was the dominant
mechanism responsible for the contrast reduction.

The spatial coherence of the atom beam is determined by the
spatial extent of the atom beam source, and describes the
correlation of different transverse points of the wave function.
When $\Delta L=1$ mm there is about 60 nm of shear (i.e.
transverse displacement) between the two interfering beams, since
$\theta=\lambda_{dB}d^{-1}\approx 60\ \mu$rad.  As a result the
contrast will be reduced because the regions of overlap for the
two interfering wave functions will be less correlated when there
is nonzero shear. The van Cittert-Zernike theorem states that the
contrast will be reduced in a way that is related to the spatial
Fourier transform of the atom beam source profile \cite{born99}.
If the source intensity distribution is assumed to be
$w_{c}^{-1}\mbox{rect}(xw_{c}^{-1})$ then the contrast will be
reduced according to\be
\begin{split}
C=C_{o}\left|\mbox{sinc}\left(\frac{w_{c}s}{\lambda_{dB}z_{c}}\right)\right|=C_{o}\left|\mbox{sinc}\left(\frac{w_{c}\Delta
L}{dz_{c}}\right)\right|,
\end{split}
\label{eq:cspat}\ee where $w_{c}$ is collimation slit width,
$z_{c}$ is distance from the collimation slit to $G_{3}$, and
$s=\Delta L\theta$ is the amount of induced shear. Equation
\ref{eq:cspat} is similar to one found in \cite{cham99} derived
using a different method.  The best fit to the data in Fig.
\ref{fig:cvsdl} using Eqn. \ref{eq:cspat} yielded a value of
$\Delta L_{01}=218\pm 72\ \mu$m.  Since $d=100$ nm and $z_{c}=2.8$
m, a best fit value of $w_{c}=81\ \mu$m was found to be consistent
with the data, according to Eqn. \ref{eq:cspat}.

Given the data in Figs. \ref{fig:phivsdl} and \ref{fig:cvsdl} the
best fit value of $\Delta L_{01}$ implies that
$\Phi_{0}-\Phi_{1}=0.39\pm 0.13$ rad, according to Eqn.
\ref{eq:phidl}. The diffraction phase $\Phi_{0}=0.30\pm 0.15$ rad
was measured recently in \cite{pifm05}, which yields the first
order diffraction phase $\Phi_{1}=-0.09\pm 0.20$ rad.  This value
for $\Phi_{1}$ is consistent with the value predicted by Eqn.
\ref{eq:psi}, as summarized in Table \ref{tab:phin}.

\begin{figure}
\scalebox{.5}{\includegraphics{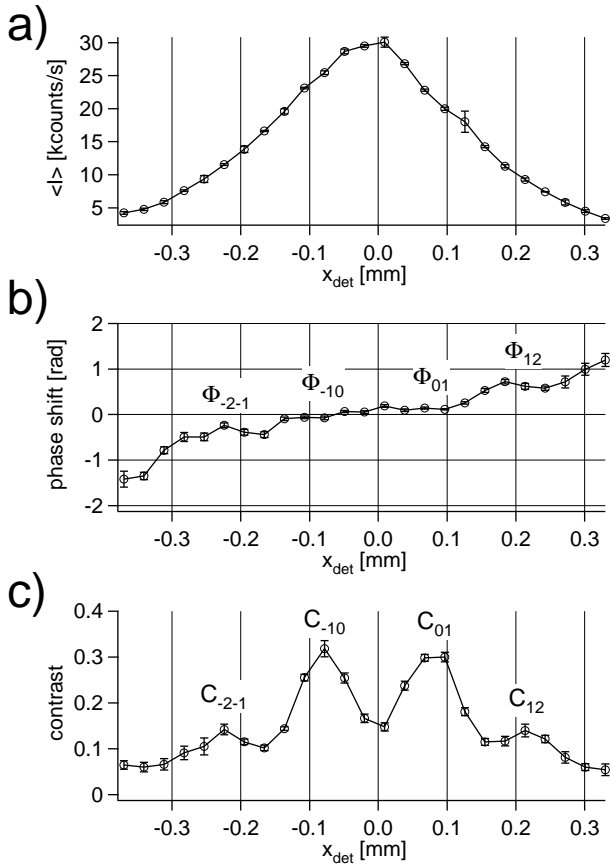}}
\caption{\label{fig:phivsxdet}Measured contrast, phase, and
average intensity as a function of detector position.  From the
contrast profile one can clearly make out the different
interferometer outputs as indicated in Fig. \ref{fig:setup}.  The
grating distance mismatch $\Delta L$ was chosen so that
$\Phi_{01}=-\Phi_{-10}\approx 0$.}
\end{figure}

By moving the detector so that it intercepts the different
interferometer outputs indicated in Fig. \ref{fig:setup} the
higher order diffraction phases can be determined.  The measured
contrast, phase, and intensity is shown in Fig.
\ref{fig:phivsxdet} as a function of detector position.  From Fig.
\ref{fig:phivsxdet} the phase $\Phi_{12}=0.4\pm 0.2$ rad can be
determined.  Equation \ref{eq:phiifm} then implies that
$\Phi_{1}-\Phi_{2}=1.57$ rad, which finally leads to
$\Phi_{2}=-1.66\pm 0.48$ rad.  This value compares well to that
predicted by Eqn. \ref{eq:psi}, as shown in Table \ref{tab:phin}.

\begin{table}[t!]
\caption{\label{tab:phin} Measured and calculated values of
$\Phi_{n}$}
\begin{ruledtabular}
\begin{tabular}{ccc}
  n & Measured $\Phi_{n}$ [rad]& Predicted$^{b}$ $\Phi_{n}$ [rad]\\  \hline
  0 & $0.30 \pm 0.15^{a}$ & $0.16$ \\
  1 & $-0.09 \pm 0.20$ & $0.02$ \\
  2 & $-1.66 \pm 0.48$ & $-1.71$ \\
\end{tabular}
\end{ruledtabular}
\noindent $^{a}$This value was measured in \cite{pifm05}.\\
\noindent $^{b}$Values are calculated with Eqn. \ref{eq:psi} and $C_{3}=2.7\ \mbox{meV nm}^{3}$ for Na atoms and a silicon nitride surface \cite{perr05}.\\
\end{table}

In conclusion the atomic diffraction phases $\Phi_{n}$ induced by
a material grating structure have been measured for $n=0,1,2$.
This was accomplished by comparing the relative phases of the
various outputs of a three-grating Mach-Zehnder atom
interferometer. These measurements agree with a simple model that
includes a vdW atom-surface interaction between the Na atom beam
and the silicon nitride grating.  In the future, one could
simultaneously monitor $\Phi_{01}$ and $\Phi_{-10}$ with two
detectors, while applying some interaction to say the $n=1$ arm.
One could then perform self-referencing phase measurements with
the interferometer, eliminating the need to take control data in a
serial fashion.  In this type of application knowledge of the
diffraction phases observed here would be required.

This work was supported by Research Corporation and the National
Science Foundation Grant No. 0354947.


\end{document}